# Manipulating Multistage Interconnection Networks Using Fundamental Arrangements

E. Gur[1] and Z. Zalevsky[2]

[1] Faculty of Engineering, Shenkar College of Eng. & Design, Ramat Gan, 52526, Israel

`gur.eran@gmail.com`

[2] School of Engineering, Bar Ilan University, Ramat Gan, 52900, Israel

`zalevsz@eng.biu.ac.il`

*Abstract*

*Optimizing interconnection networks is a prime object in switching schemes. In this work the authors present a novel approach for obtaining a required channel arrangement in a multi-stage interconnection network, using a new concept – a fundamental arrangement. The fundamental arrangement is an initial N-1 stage switch arrangement that allows obtaining any required output channel arrangement given an input arrangement, using N/2 binary switches at each stage. The paper demonstrates how a fundamental arrangement can be achieved and how, once this is done, any required arrangement may be obtained within 2(N-1) steps.*

## 1. Introduction: Multi-stage interconnection networks

One common use for multi-stage processing is in Optical switching (Saleh and Teich, 1991 and Pan et al., 1999). It has been shown that Multi-stage Interconnection Networks (MIN) are the preferred way to implement compact switches by means of shuffle and exchange (Parker, 1980). In recent years much work has been done in the field of compact optical MINs (Reinhorn et al., 1997 and Marom et al., 1998). This led the way to an all-optical switch (Cohen et al., 1998 and Mendlovic et al., 1999), which is faster and more flexible than common electronic switches used for fiber communication. The multi stage switching setup is similar to the multi stage processor and many concepts may be shared between the two.

First we define some common classes of connectivity. Partial connection: One input channel may connect to at least one output channel. Fully Connected System: Any single input can be connected to any (arbitrary) output. However, after this is achieved it may prevent other connections from being implemented. This condition is known as a BLOCK (Clos, 1953). Rearrangeable Non-blocking (RNB) System: Any permutation of an input-to-output connection can be established. However, if a new connection has to be made, some of or all the existing connections have to be reconnected, thus introducing a temporary channel interruption until the rearrangement is complete. Wide-sense Non-blocking (WSNB) System: If while connecting the inputs to the outputs a proper algorithm is applied, a new connection can be established without disturbing any of the existing ones. WSNB allows any set of input channels to connect with any set of output channels with no need to alter the dynamic routing structure except in specific modules, connected directly to the alteration. Unfortunately, in standard MIN architectures, a general solution for wide-sense non-blocking is yet to be found. Strictly non-blocking (SNB): Allows any set of input channels to connect with any set of output channels with no need to alter the dynamic routing structure except in specific modules, regardless of the inner-structure. Unfortunately, in standard MIN architectures, strictly non-blocking cannot be achieved.

To connect a set of input channels to a set of output channels one might use a crossbar switch as described in Figure 1. In this structure any input channel may be connected to a free (unconnected) output channel without any blocking, always. It also allows broadcasting a single channel to several output channels in a simple manner. However, such a structure requires $N^2$ switches, which is a large number of dynamic elements.

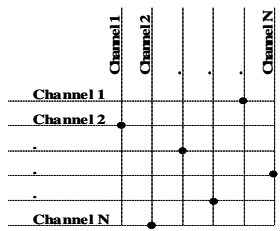

Figure 1. An N-by-N crossbar switch connecting N lines to N lines.

For this reason we turn to Multi-stage Interconnection Networks (MIN). In these networks part of the dynamic switching is replaced by static shuffling. At each stage a set of switches can convert the input channel arrangement into one of a finite number of arrangements. For obtaining the ability to allow any channel arrangement at the output, one requires several switching stages accompanied by shuffling/routing stages to ensure that each switching stage has the opportunity to deal with different channel combinations.

Now we define some common MIN routing schemes: Perfect shuffle, Banyan and Crossover. The best way to define these routing/shuffling methods is demonstrated in Figure 2:

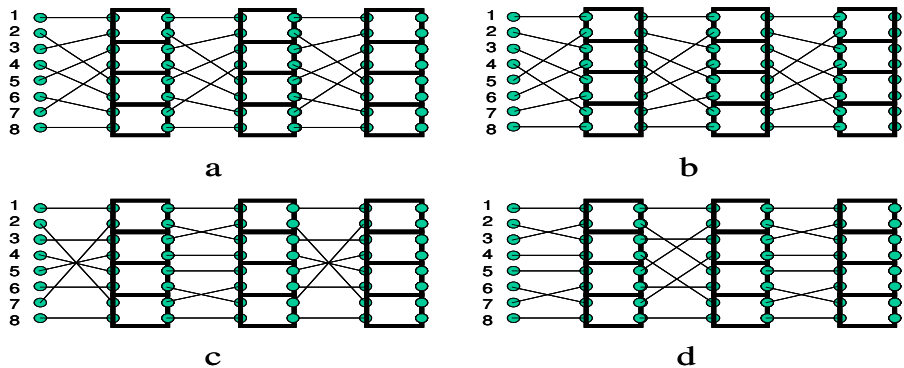

Figure 2. The first 3 stages of an 8 by 8 MIN setup using the following routing methods (a) Perfect Shuffle, (b) reverse Perfect Shuffle, (c) Crossover, and (d) Banyan.

In perfect shuffle we shuffle all channels the same way we may shuffle a deck of cards, by splitting the deck in two and then taking cards alternately from the two decks. In this routing scheme all routing stages are identical, and the same shuffling applies to all stages. A reverse shuffle is also commonly used.

In Crossover, the first stage deals with the entire set of channels and implements a certain shuffle. The second stage deals with half the channels using the same shuffling concept and so on, until the minimal shuffle is achieved (crossing or transmitting between channels sharing the same switch is not considered shuffling) and then returning to the initial shuffle, addressing all channels.

In Banyan, a slightly different shuffling concept is chosen, and this time the number of channels involved in the shuffling increases as we advance to the next stage. Once all the channels are used in a single shuffle we return to the initial shuffle, as in the first stage.

Here we find it necessary to define topological equivalence: Two networks A and B are said to be topologically equivalent (or isomorphic) when the links and switches in network A can be relabeled with logical addresses so that the resulting topological connections in network A are identical to the topological connections described by network B. It is needless to state that all the architectures shown in Figure 2 are topologically equivalent.

In 1986, a simple optical realization of perfect shuffle was suggested (Lohmann et al., 1986). A simplified version of this setup is shown in Figure 3 and demonstrates a perfect shuffle for eight channels using two lenses and two micro-prism arrays.

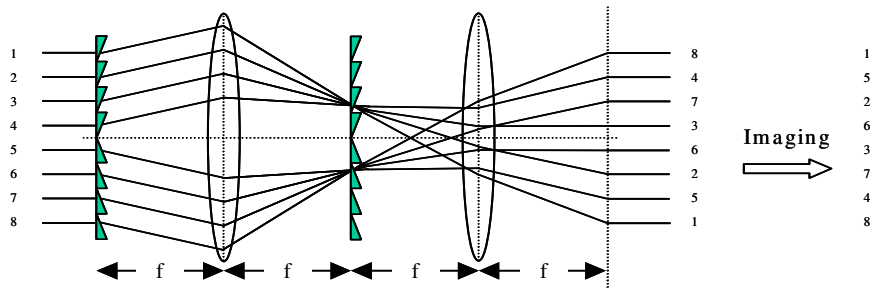

Figure 3. An optical realization of perfect shuffle routing for eight channels.

The micro-prisms are the diffractive equivalent of full size prisms, allowing a slim design for the setup and enabling its implementation using planar optics. At the same time, Lohmann published a more general work on the possible implementations of digital optical computers (Lohmann, 1986). These possibilities, alongside the novel optical switches developed in the 1990s allow the construction of complete MINs using optics.

We will focus on the Omega-2 ($\Omega$-2) network that uses dual-channel switches (as illustrated before). In such cases the switches may occupy one of the following states: bypass, and exchange. Using such switches in an MIN setup generates a sufficient flexibility to implement RNB networks, assuming the number of stages is large enough. A more detailed representation of shuffling and switching will be given in the specified section, when the authors makes an effort to model the MIN and to find solutions to specific problems.

## 2. Using the iterative procedure in MIN optimization

First we recall that the MIN can be optimized in an iterative manner (Gur et al., 2002). We start with an input vector, where each element in the vector corresponds to a single input channel. Our purpose is to manipulate these input channels in such a way that the output vector will contain the same channels, yet in a different, required, order. Each routing stage contains static, or fixed, shuffling architecture (Golomb, 1961, Parker, 1980) such as perfect shuffle, Banyan or crossover, given in Figure 4. We present the matrix form of the shuffling architectures since we use these matrix notations later on. Each stage of static routing is followed by an active routing stage, composed from bypass/exchange switches. We recall that a bypass/exchange module accepts two input channels and either leaves them as they are (bypass) or switches between them (exchange). In this way, one might reach any desired output from any given input if the number of stages is large enough. The matrix representation of these switches is shown in Figure 4.

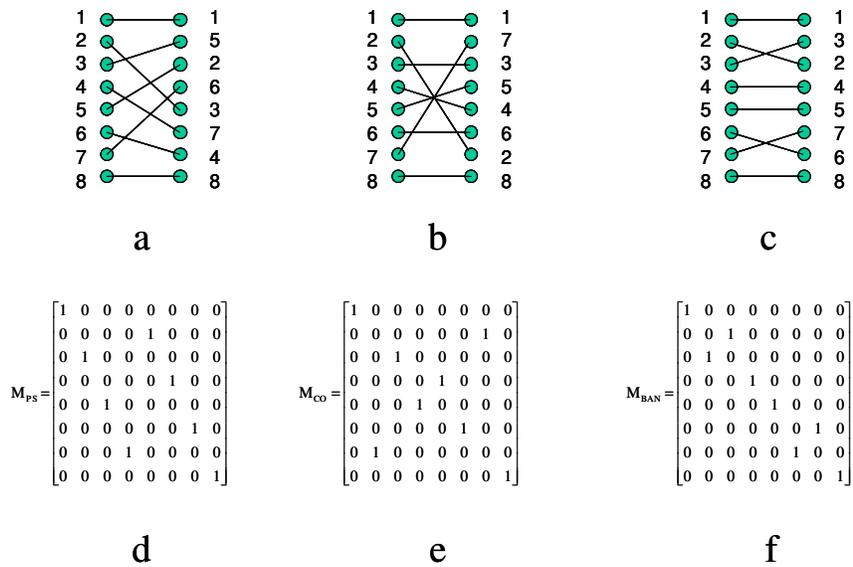

Figure 4. The first stage in shuffling architectures (a) perfect shuffle, (b) crossover and (c) Banyan, all for an 8 by 8 setup. In (d) through (f) one may view the matrix implementation of the routing given in (a) through (c), respectively. In perfect shuffle, all stages are identical, in contrary to crossover and Banyan.

Given an input vector and an output vector (each representing a set of channels), there are several matrices representing a linear system that will lead from one to the other. However, there are a few constraints on the matrices available for switching. The dynamic routing matrix must contain only "0"s and "1"s and no more than one "1" in a given row or column (each input channel can be connected only to one output channel and each output channel can receive information only from one input channel at a time). A logical "1" in a given row and column coordinate means that the input channel corresponding to the column index is connected to the output channel corresponding to the row index. The matrix must be constructed out of bypass 2 by 2 matrices and exchange 2 by 2 matrices (as drawn in Figure 4), placed on the main diagonal, where the other elements of the matrix are all "0". An example for valid and non-valid bypass/exchange matrices is given Figure 5.

The output of the entire MIN is a result of M stages of both static and dynamic routing (the elements in the input and output vectors are the indices of the input channels). In the input vector the element location is identical to its value whereas in the output vector an index i in the j location refers to the i$^{th}$ input channel moved to the j$^{th}$ output channel). The path of proceeding from input to output is as follows. First, the input vector is multiplied by a static routing matrix (in the perfect-shuffle architecture all routing matrices are identical, in other architectures, such as Banyan and crossover architectures, each stage the routing matrix combines a different number of participants, as shown before). Then, the result is multiplied by a switching (dynamic routing) matrix, then by a second static routing matrix and so on. If the number of stages is large enough (an order of N or even $\log_2 N$ stages, depending on the value of N) one can obtain any output vector from any input vector.

$$bypass\,sub\,matrix: \begin{bmatrix} 1 & 0 \\ 0 & 1 \end{bmatrix} \qquad exchange\,sub\,matrix: \begin{bmatrix} 0 & 1 \\ 1 & 0 \end{bmatrix}$$

$$\begin{bmatrix} 1 & 0 \\ 0 & 1 \end{bmatrix} \begin{bmatrix} c1 \\ c2 \end{bmatrix} = \begin{bmatrix} c1 \\ c2 \end{bmatrix} \qquad \begin{bmatrix} 0 & 1 \\ 1 & 0 \end{bmatrix} \begin{bmatrix} c1 \\ c2 \end{bmatrix} = \begin{bmatrix} c2 \\ c1 \end{bmatrix}$$

Figure 5. Bypass and exchange sub matrices. The bypass matrix leaves the inputs as they are while the exchange matrix switches between the two channels.

## 3. Different connectivity architectures– discussion

In this section we demonstrate the difference between Wide-Sense Non-Blocking (WSNB) and Strictly Non-Blocking (SNB) and we try to define a general condition for making Wide-Sense Non-Blocking possible in common MIN. We also suggest a method for constructing an arrangement of switches, referred to as a "Fundamental Arrangement" (FA) that contains all possible pairs of channels in an N-1 stage network, and can be used as a basis for fast Rearrangeable Non-Blocking (RNB) MIN. We do this with respect to the perfect shuffle arrangement, yet without restricting generality since all MIN architectures are topologically equivalent and thus we can find a similar FA for other architectures as well.

### 3.1 SNB & WSNB – definitions and existence in omega-2 MIN

First we recall the definition of a Wide-Sense Non-Blocking network: If while connecting the inputs to the outputs a proper algorithm is applied, new connection can be established without disturbing any of the existing ones. If we take a closer look at the underlined condition and try to interpret it in terms of an omega-2 MIN, we find that it refers to the case where switching any two channel may be done using a single 2 by 2 switch, thus not disturbing the other channels. For this to be true we need that at every moment the MIN will contain all possible switches. This might seem logical if we do not limit the number of stages of the MIN. However, as we will show below, it requires a certain condition. We first look at a multi stage MIN setup as described in Figure 6.

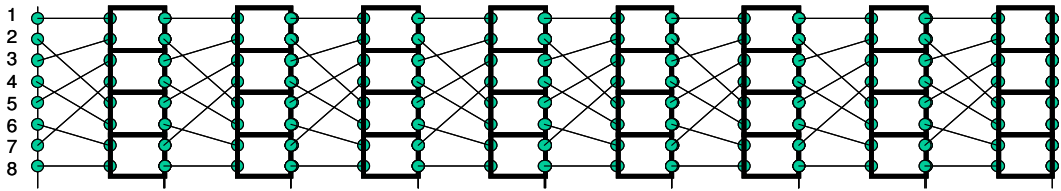

Figure 6. An 8 by 8 MIN setup using an omega-2 structure and perfect shuffle routing.

Although we look at the perfect shuffle architecture, the following remarks apply to all topologically equivalent architectures. Let us assume that SNB is possible without restrictions, and try to find the requirements of the WSNB algorithm. Therefore, if the number of stages presented is sufficient, we will observe all possible switch input combinations. Therefore, any input-output connection may be obtained by changing a single switch, without disturbing other connections, and performing this change will leave the system Wide-Sense Non-Blocking (and SNB if we do not force any demands). Now we start changing the switches one at a time. For the prefect shuffle scheme we simply convert each exchange to a bypass. Every change must leave the system SNB; otherwise SNB has no meaning. Finally we obtain an all-bypass network. Now we note that due to the fully connected system, after $\log_2 N$ stages (N being the number of channels), the all-bypass MIN returns to its original channel arrangement, as shown in Figure 7. For a different shuffling scheme the switch combination allowing this characteristic is different, of course.

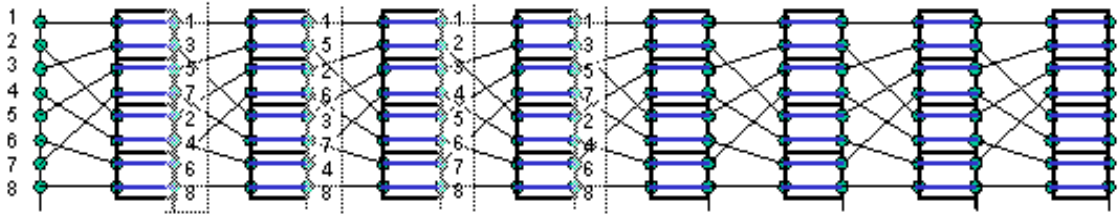

Figure 7. The 8 by 8 MIN setup when all switches are in bypass mode.

Since after $\log_2 N$ stages we return to the original channel arrangement, each channel encounters only $\log_2 N$ channel along its path, and it encounters them over and over again. Thus, the other $N-1-\log_2 N$ channels are not interacting directly with the specific channel and therefore not all pairs are represented by independent switches. The direct conclusion is of course that SNB does not exist in MIN, but this is a known fact.

The new distinction is that a block of consecutive $\log_2 N$ stages of all-bypass switches is the main reason for preventing SNB and therefore the key for planning WSNB. In the early 80s Benes and Kurshan (Benes & Kurshan, 1981) showed the private case of a 4-by-4 MIN that requires 4 stages for WSNB with the accompanied algorithm that disallows the 2 middle stages from being all-bypass. This is merely a private case of our conclusion with respect to $\log_2 N = 2$ for the case $N = 4$. Over the years many attempts have been made to find a general solution to the WSNB problem, yet different solutions were suggested for different N values. We hope that the above notation might lead the way to a more general WSNB algorithm. The general construction of WSNB is quite difficult and therefore we choose to concentrate on RNB as in the next section.

If we insist on an SNB setup we can obtain it in non-MIN architectures such as the one shown in Figure 8, where $N^2$ direct connections are required.

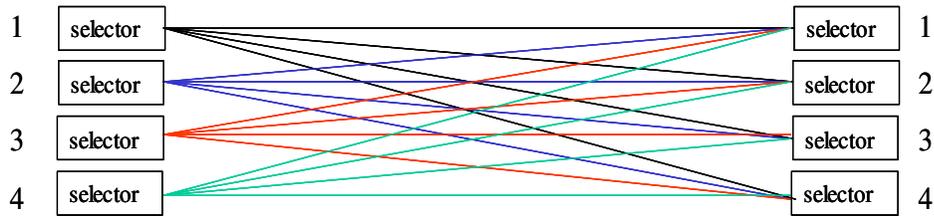

Figure 8. A Strictly Non-Blocking architecture allowing to switch any pair of channels without disturbing the other channels.

### 3.2 RNB and the construction of fundamental arrangements

As shown before the existence of all switching pairs at a specific arrangement does not indicate the feasibility of WSNB since exchanging the channels for a single switch may eliminate several of the other pairs. However, the existence of such an arrangement, we will refer to as a Fundamental Arrangement (FA), does hold certain characteristics regarding the system connectivity.

In recent years a considerable amount of research related to RNB networks in general (Shen at al., 2002) and to RNB in MIN in particular (Das et al., 2000 and Cam, 2003) was conducted. In the later research it was shown that $(2\log_2 N - 1)$ stages are sufficient for obtaining RNB in MIN architecture. This is known to be the theoretical lower bound for RNB. However, the algorithm used for rearranging a general set of channels is quite complex. It also relates, as most discussion, to a specific number of stages N. In the following paragraphs we present a simple algorithm applicable for large MIN architectures ($N-1$ stages).

We look again at the perfect shuffle routing scheme, without restriction of generality. Since the structure is symmetrical with respect to the y-axis in order not to return to a pair of channels occupying a single switch before passing through all other pair combinations we use an anti-symmetric switch choice as the one given in Figure 9. As seen from the MIN sketch, choosing a non-symmetric structure, and one that differs from one stage to the next we can obtain an FA. This concept was tested for systems containing 4, 8, 16, 32 and 64 channels and it produced the FA in each and every one.

We note that we need only $(N-1)$ stages to obtain an FA since we have $N/2$ switches per stage and the total number of pairs is $(N-1)N/2$. This is since the first channel interacts with $(N-1)$ channels, the second with $(N-2)$ channels excluding the first, etc.

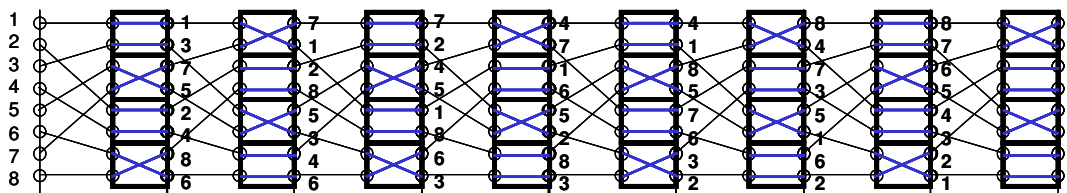

Figure 9. Obtaining a fundamental arrangement in an 8-channel MIN using perfect shuffle.

We will now show that from the FA one can reach any output combination, thus that $(N-1)$ stages are sufficient for obtaining RNB. We start with the 8-by-8 case and explain the procedure via an example, and then we give a more general explanation.

Let us assume that we have the 7 stages setup given in Figure 9 and we want the output to become (from top to bottom): 1,3,5,7,2,4,6,8. Thus, from comparison to the current FA output, we look for the following channel pairs: 8-1, 7-3, 6-5, 5-7, 4-2, 3-4, 2-6 and 1-8. We start from the 1st stage and find the pairs: 5-7 and 2-4. We switch them making the output 8,5,6,7,2,3,4,1. So now we require different pairs: 8-1, 5-3, 6-5, 3-4, 4-6 and 1-8. None of them appears in the 1st stage, nor in the 2nd stage now being 5,1,4,8,7,3,2,6. We proceed to the 3rd stage where we find 1-8. We change this switch and obtain at the output: 1,5,6,7,2,3,4,8.

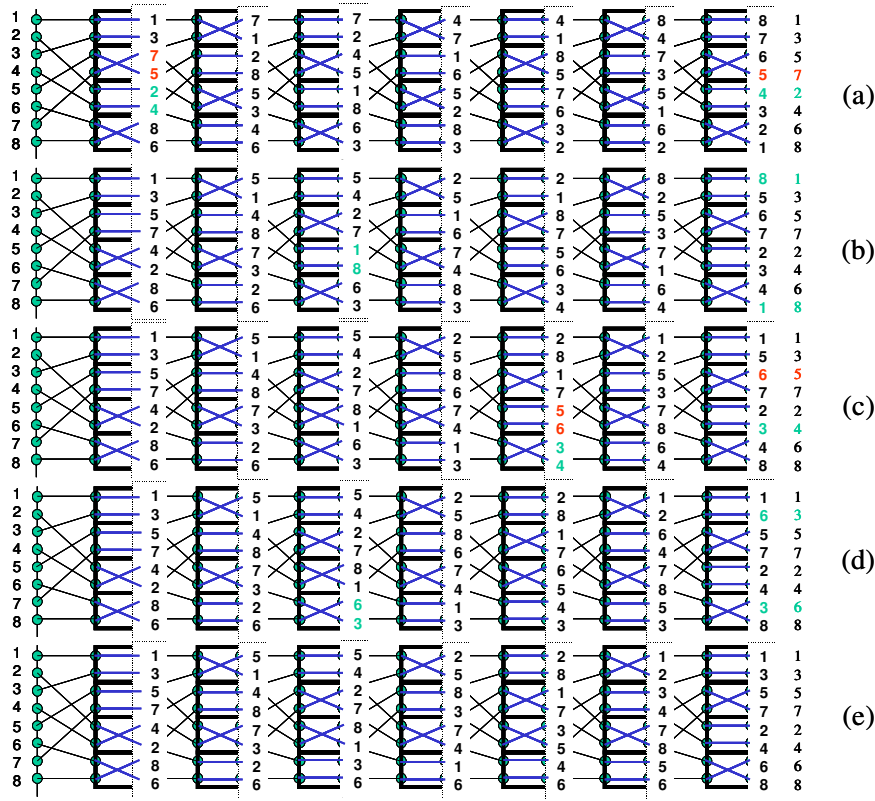

Figure 10. A 5-step example for obtaining a required output from the basic FA output.

Next we need to change 5-6 and 3-4, both present at the 5th stage. The output now is 1,6,5,7,2,4,3,8 and thus we need only to find the pair 3-6, and it exists in the 3rd stage. Thus, after altering the last switch, we obtained the required output from the FA arrangement. From this example we anticipate that there's a way to obtain any output arrangement by first changing the switches to the FA and then proceeding according to the steps described above. The network described in this example is shown in Figure 10. For a more general description we present the next paragraphs.

We start with $(N-1)N/2$ switches, according to FA, such that all couples are present. Let us give arbitrary names to the input channels ('1' does not have to be the actual first channel etc.) and use the 8-channel case for better understanding. The alterations required to switch from one output order to another has a variety of options from 4 cyclic groups of 2 elements: 1-2, 3-4, 5-6, 7-8, meaning that 4 switches are enough and there's no cross-connection between pairs, to the complete cycle (remember the names of the channels are arbitrary) which means actually no cycle: 1-2-3-4-5-6-7-8 (which means channel 1 is redirected to the second position, channel 2 to the third position etc., leaving channel 8 to occupy the first position), through various arrangements such as: 1-2-3-1, 4-5, 6-7-8-6.

Without restriction of generality we write the following:
- At a given stage channels 1 and 2 where swapped using a single switch. This switch will be denoted as S12 (In the worst case scenario this will happen in the first step, using the FA).
- Thus, switches in following stages have changed if one of their inputs was 1 or 2, e.g., the previous S15 will become S25.
- Next, a new pair of channels has to be swapped, and the options are as follows:

1. The new pair contains channels such as 3 and 4, i.e., no connection with S12. Therefore, it is certain that S34 exists, and a single switch may be used.
2. The new pair contains channels such as 1 and 3 (or 2 and 3), and thus there are three sub-options:
    2.1. In the FA, S13 preceded S12 (comes at an earlier stage) and therefore it still exists. Hence, channels 1 and 3 may be swapped using a single switch.
    2.2. In the FA, S13 follows S12 (comes at a later stage), but S23 also follows S12. Thus, S13 and S23 will simply change roles, i.e., S13 exists and channels 1 and 3 may be swapped using a single switch.
    2.3. In the FA, S13 follows S12, but S23 precedes S12. This means that after S12 has been switched S13 becomes S23. Thus, two switches S23 are present, one before S12 and one after S12. The later S23 may be used to generate an indirect connection between channels 1 and 3, which may require concatenation (chaining) of several switches. The earlier S23 will be used to return to the initial state, except for swapping 1 and 3.

In 2.1, it has been noted that: "In the FA, S13 preceded S12 (comes at an earlier stage) and therefore it still exists. Hence, channels 1 and 3 may be swapped using a single switch". We note that changing S13 will change S12 and therefore it may vanish, BUT we will use S13 only if S12 was used to place channel 2 in the proper position. If however, S12 was used to place channel 1 in the proper position, then the next switch we look for will be S2#, e.g. S24, which involves channel 2 rather than channel 1.

Finally we write the straightforward algorithm as follows:
1. Place the switches in the FA arrangement.
2. Compare the initial output to the required output and thus obtain N required pairs switches (or less if a certain channel is already in place).
3. Search for any of the pairs in the first stage and move up the stages until a pair (or more) is found in the Kth stage. Such a switch is sure to be found within the N-1 stages.
4. Change the switch position from bypass to exchange or vice versa.
5. Obtain the current output and compare it to the required one to obtain the new required pairs (less than N since at least one channel is now in place).
6. Search for any of the pairs in the first stage and move up the stages until a pair (or more) is found.
7. Proceed with steps 4 to 6 until all output channel are in place.

It is important to note that due to the FA we will never need more than (N-1) stages since the first switch may eliminate up to (N-1) switches, the second up to an extra (N-2) switches (since combinations of pairs from both steps are not eliminated) etc., until finally the Nth switch eliminates no switches. Summing up we obtain $(N-1)N/2$ different switches as present initially in the FA. Comparable research either showed an increase in the number of stages by a factor of 3 (Bergen, 2003), or used iterative methods (Khandker et al., 2002) thus requiring a longer time to converge to the final arrangement.

Thus, we presented an algorithm for obtaining any required channel arrangement from any given arrangement using a maximum of 2(N-1) steps: first we find the required steps from the FA to the original arrangement and reverse the steps to obtain the FA, next we find the required steps from the FA to the required channel arrangement. Each of these two parts contains a maximum number of (N-1) steps and thus the entire procedure requires 2(N-1) steps at the most. It is clear that this is not always the fastest way to obtain one channel arrangement from to another (e.g., if they differ by a single switchable pair) but it is a no-fail method.

## 4. Conclusions

In this paper the authors showed a novel algorithm required to obtain RNB connectivity using analytical procedures rather than iterative procedures. The suggested approach uses a reference arrangement named a fundamental arrangement. The original channel arrangement is first transformed into a fundamental arrangement in N-1 steps, at the most, and then transformed into the required channel arrangement. The result is a fast routing procedure that uses only N-1 stages and 2(N-1) steps to obtain the required output, in contrary to an order of $N \log_2 N$ stages used in previous algorithms. Finally, the results led to a discussion on several insights regarding WSNB and SNB architectures. Although the method is shown for the perfect shuffle arrangement, all MIN architectures are topologically equivalent and therefore the results apply to any multi-stage architecture.